\pdfoutput=1
\documentclass[%
 reprint,
 amsmath,amssymb,
 aps,
]{revtex4-2}

\usepackage{amssymb,amsmath,epsfig,latexsym,booktabs,braket,
gensymb,amsbsy,stackrel,slashed}

\usepackage[dvipsnames]{xcolor}
\usepackage{tikz}
\usepackage{pgfplots}
\usetikzlibrary{patterns}
\usepackage[autooneside=false]{scrlayer-scrpage}

\usepackage{amsfonts}
\usepackage{amssymb}
\usepackage{amsmath}
\usepackage{mathtools}

\usetikzlibrary{
  arrows,
  backgrounds,
  calc,
  decorations.markings,
  decorations.pathmorphing,
  decorations.pathreplacing,
  fit,
  intersections,
  mindmap,
  patterns,
  positioning,
  through}

\usepackage{mathtools}
\usepackage{amstext}
\usepackage{amsfonts}
\DeclareSymbolFontAlphabet{\amsmathbb}{AMSb}
\usepackage{physics,braket,bbold}

\usepackage{graphicx}
 \usepackage[all,cmtip]{xy}
\usepackage{dcolumn}
\usepackage{bm}

\newcommand{\cN}{{\cal N}}
\newcommand{\cL}{{\cal L}}
\newcommand{\nFour}{$\cN \, = \, 4$ }
\newcommand{\rar}{\, \rightarrow \,}
\newcommand{\cO}{{\cal O}}
\newcommand{\eqsp}{\, = \,}
\newcommand{\la}{\langle}
\newcommand{\ra}{\rangle}

\newcommand{\cD}{{\cal D}}
\newcommand{\cK}{{\cal K}}
\newcommand{\beq}{\begin{equation}}
\newcommand{\eeq}{\end{equation}}

\newcommand{\mh}{\mathfrak{h}}
\newcommand{\ep}{\epsilon}
\newcommand{\cZ}{{\cal Z}}

\usepackage{color}

\begin{document}

\hfill \small{HU-EP-23/54}

\title{Anomalous dimensions from the \nFour SYM hexagon}

\author{Burkhard Eden}\email{eden@math.hu-berlin.de}
\affiliation{%
Institut f\"ur Mathematik und Physik, Humboldt-Universit\"at zu Berlin, Zum gro{\ss}en Windkanal 2, 12489 Berlin, Germany.}

\author{Maximilian Gottwald}\email{gottwalm@physik.hu-berlin.de}
\affiliation{%
Institut f\"ur Mathematik und Physik, Humboldt-Universit\"at zu Berlin, Zum gro{\ss}en Windkanal 2, 12489 Berlin, Germany.}

\author{Dennis le Plat}\email{diplat@physik.hu-berlin.de}
\affiliation{%
Institut f\"ur Mathematik und Physik, Humboldt-Universit\"at zu Berlin, Zum gro{\ss}en Windkanal 2, 12489 Berlin, Germany.}

\author{Tobias Scherdin}\email{scherdit@physik.hu-berlin.de}
\affiliation{%
Institut f\"ur Mathematik und Physik, Humboldt-Universit\"at zu Berlin, Zum gro{\ss}en Windkanal 2, 12489 Berlin, Germany.}

\date{\today}

\begin{abstract}
We consider the correlator $\la \cL \, \cK \, \tilde \cK \ra$ of the Lagrange operator of \nFour super Yang-Mills theory and two conjugate two-excitation operators in an $su(2)$ sector. We recover the planar one-loop anomalous dimension of the renormalised operators from this hexagon computation.

\end{abstract}

\maketitle


\section{Introduction}

The on-shell Lagrangian density of \nFour super Yang-Mills theory (SYM) can be written as \cite{Eden:2011yp}
\beq
\begin{aligned}
\cL \eqsp \frac{1}{g_\mathrm{YM}^2} \tr\bigg(- \frac{1}{2} &F_{\alpha \beta} F^{\alpha \beta} + \sqrt{2} \, \Psi^{\alpha I} [\Phi_{IJ}, \Psi^J_\alpha] \, \\
&- \, \frac{1}{8} \, [\Phi^{IJ},\Phi^{KL}][\Phi_{IJ},\Phi_{KL}] \bigg) \, , \label{ourL}
\end{aligned}
\eeq
which contains the field strength $F^{\alpha \beta}$, four fermions $\Psi^{\alpha I}$ and the antisymmetric scalar fields $\Phi^{IJ}$, where $I,J,K,L \, \in \{1 \ldots 4\}$.

Correlation functions as defined by the path integral
\beq
\la \cO_1 \ldots \cO_n \ra \eqsp \int \cD(\Phi, \Psi, \bar \Psi, A) \, \cO_1 \ldots \cO_n \, e^{ i \int d^4x_0 \, \cL_0 } 
\eeq
should therefore obey the identity
\beq
g^2 \frac{\mathrm{d}}{\mathrm{d}(g^2)} \, \la \cO_1 \ldots \cO_n \ra \eqsp \frac{-i}{g^2} \int d^4x_0 \, \la \cL_0 \, \cO_1 \ldots \cO_n \ra \, . \label{intri}
\eeq
Initially intended as a criticism \cite{Intriligator:1998ig}, this observation has been instrumental in constructing correlators of half-BPS operators at the integrated level at one- and two-loop order, and as integrands to very high order in the coupling $g_\mathrm{YM}$ \cite{Howe:1999hz,Eden:2000mv,Eden:2012tu,Bargheer:2022sfd}.

In the present work we want to study equation \eqref{intri} as a relation between the two-point function of a renormalised scalar primary operator $\mathcal{K}$ and its conjugate $\tilde{\mathcal{K}}$, and the three-point function obtained by inserting $\mathcal{L}$ into it. At \emph{Born level}
\beq
\la \cL_0 \, \cK_1 \, \tilde \cK_2 \ra \eqsp \frac{c(N)}{{(x_{01}^2)}^2 {(x_{02}^2)}^2 {(x_{12}^2)}^{\Delta_0-2}} \, .
\eeq
Configuration space Feynman integrals yield the functional dependence stated in the last formula. Here $\Delta_0$ is the naive scaling dimension of $\cK$, and so the computation serves to determine the constant $c(N)$, where $N$ is the rank of the gauge group.

The space time integral over the insertion point 
\beq
\int \frac{d^4x_0}{x_{01}^4 x_{02}^4} \nonumber
\eeq
is divergent in this situation. 
It reproduces the one-loop divergence of the two-point function on the left of \eqref{intri} so that upon regularising we must have
\beq
c(N) \, \propto \, \gamma_1\, \label{gamRel}
\eeq
if the planar scaling dimension of $\cK$ has the coupling constant expansion $\Delta_{\cK} \eqsp \Delta_0 + g^2 \gamma_1 + g^4 \gamma_2 + \ldots$ in terms of the 't Hooft coupling $g^2 \eqsp g^2_\mathrm{YM} \, N / (8 \, \pi^2) \, .$ \\

The spectrum problem of \nFour SYM 
is related to finding the energy eigenstates of the Heisenberg spin chain \cite{Minahan:2002ve,Beisert:2004hm}. More recently, structure constants became accessible to such \emph{integrability} methods via the hexagon formalism of \cite{Basso:2015zoa}. In this article we present a first computation verifying \eqref{gamRel} entirely within the Bethe ansatz framework.

In the original spin chain approach to \nFour theory the scalar $Z = \Phi^{34}$ is regarded as a vacuum and other fields can travel over a chain of such sites \cite{Minahan:2002ve}. Unfortunately for our endeavour, the complex conjugate vacuum $\bar{Z}= \Phi^{12}$, half of the fermions, and the field strength $F^{\alpha \beta}, \dot{F}^{\dot \alpha \dot \beta}$ are excluded from the set of excitations, and hence it would appear that none of the terms of the on-shell Lagrangian \eqref{ourL} is captured.

On the other hand, for the $so(6)$ sector comprising all scalar fields it has been known for a long time how to realise the missing conjugate of the vacuum as a \emph{double excitation}, by placing two scalars on the same site. 
The missing fermions and the two parts of the field strength tensor are provided by further types of double excitations \cite{Eden:2023ygu}.

Customarily, Bethe equations are diagonalised by the \emph{nested Bethe ansatz} \cite{PhysRevLett.19.1312}. Having found a solution, its primary roots can be used also in the original Bethe ansatz for the \nFour spectral problem, in which the (restricted set of) excitations scatter on the chain by a $psu(2|2)$-invariant $S$ matrix \cite{Beisert:2005tm}. Operators in higher-rank sectors are then described by a multi-component wave function. 

The operator \eqref{ourL} is a supersymmetry descendent of the half-BPS operator $\cO^\mathbf{20}$. Acting with all four generators $Q_{a'}^\alpha$ yields
\beq
\cL \, = \, \frac{1}{4} Q_3^1 \, Q_4^1 \, Q_3^2 \, Q_4^2 \, \tr(\Phi^{34} \Phi^{34}) \, . \label{q4L}
\eeq
These generators transform the spin chain vacuum as
\beq
Q_3^\alpha \, \Phi^{34} \eqsp \Psi^{\alpha 4} \, , \qquad Q_4^{\alpha} \, \Phi^{34} \eqsp - \Psi^{\alpha 3} 
\eeq
allowing us to identify the Lagrange operator with the four excitations \cite{Eden:2023ygu}
\beq
\{ \Psi^{13}, \Psi^{23}, \Psi^{14}, \Psi^{24} \} \, \label{myFour}
\eeq
on a length two chain.
At lowest order in the coupling the Bethe wave function exists because the fermions may occupy the two sites in pairs
\beq
-Q_3^\alpha \, Q_4^\beta \, \Phi^{34} \eqsp \ket{\substack{\ \Psi^{\alpha 4} \\ \Psi^{\beta 3} }} \eqsp F^{\alpha \beta} \, .
\eeq
The stacked state in the middle denotes a double excitation at one site. On the other hand, the level one wave function on which the four excitations scatter with the $S$ operator of \cite{Beisert:2005tm} dissolves the concept of spin chain length.
However, the local structure of the wave function is of no further relevance in spectrum computations. Even if the length two Bethe solution with four excitations only exists owing to double excitations, the latter will not play any further r\^ole in calculations.
Moreover, the hexagon \cite{Basso:2015zoa} does also not depend on the local wave function.
In \cite{Eden:2022ipm,Eden:2023ygu} it is illustrated on a series of examples that it does correctly compute structure constants for operators requiring the presence of double excitations.

These calculations necessarily involve multi-component wave functions. A priori, we have to write one wave function for every initial ordering of the four fermions in \eqref{myFour} and of the cases
\beq
\{\Psi^{13}, \Psi^{13}, \Psi^{24}, \Psi^{24} \}, \, \{\Psi^{23}, \Psi^{23}, \Psi^{14}, \Psi^{14} \} \, . \label{degenerate}
\eeq
Each of the resulting 36 wave functions comes with a coefficient. Off shell, these amplitudes are uniquely determined matching the entire state on the nested Bethe ansatz \cite{Eden:2022ipm}\footnote{If the full coupling constant dependence is supposed to be incorporated, only selected gradings of the nested Bethe ansatz can be chosen \cite{Beisert:2005fw}. In the \emph{Born approximation} relevant to our purposes any grading works.}.

On shell, the situation may be degenerate for descendent states, ie. when there are infinite Bethe rapidities. One of our aims is to decide whether all 36 wave functions are necessary to recover \eqref{gamRel}. Our answer will be no: on a gauge invariant state, the four distinct supersymmetry generators in \eqref{q4L} anticommute. Consequently, the sequence of taking the supersymmetry variations can alter the result only by an overall sign, and for the choices in \eqref{degenerate} the variation vanishes. Excitingly, the hexagon computation presented below has the very same features: the initial ordering of the four magnons only results in an overall sign, and the 12 cases from \eqref{degenerate} rather non-trivially yield zero.

\section{The computation}

The $su(2)$ sector of the \nFour spin chain model has $n$  magnons $X \eqsp \Phi^{24}$ travelling over a closed chain of $L$ sites, so there are $L-n$ further fields $Z \eqsp \Phi^{34}$. Each magnon moves with a quasi momentum $p_j$, or equivalently the \emph{rapidity}
\beq
u_j \eqsp \frac{1}{2} \cot \frac{p_j}{2} \, .
\eeq
In the planar approximation, the one-loop conformal eigenstates for every $L$ are exactly given by the wave functions of the coordinate Bethe ansatz. Here we are chiefly interested in their eigenenergies viz anomalous dimensions. These can be found from the \emph{Bethe equations}
\begin{equation}
e^{i \, p_j \, L} \, \prod_{k \neq j} S_{jk} \, = \, 1 \, , \qquad S_{jk} = \frac{u_j - u_k - i}{u_j - u_k + i} \label{eipS} 
\end{equation}
built from the shift operator $e^{i \, p_j} \eqsp (u_j + i/2)/(u_j - i/2)$ and the \emph{scattering matrix} $S$. Importantly, there is \emph{factorised scattering}: multi-particle scattering factorises into two-particle processes. Finally, translation invariance along the chain implies the zero momentum constraint $\sum_j p_j \, = \, 0$.

A set $\{u_j\}$ solving the Bethe equations are called \emph{Bethe roots}. In terms of these rapidities
the energy of a Bethe state is
\begin{equation}
\gamma_1 = \sum_{j=1}^n \frac{1}{u_j^2 + \frac{1}{4}} \, .
\end{equation}
Specialising to the two-excitation case, the zero momentum constraint yields $u_1 \eqsp - u_2$ and therefore the remaining Bethe equation simplifies to
\begin{equation}
e^{i \, p_j (L-1)} \, = \, 1 \, .
\end{equation}
The lowest two-excitation states are then characterised by the roots and energy eigenvalues given in Tab. \ref{tab:Tab1}.
\begin{table}[h]
\begin{center}
\caption{Bethe roots $u_2=-u_1$ and energy eigenvalues $\gamma_1$ for $su(2)$ primary operators of lengths $L=4,\dots,9$.} \vskip 2mm
\label{tab:Tab1}
\begin{tabular}{c|c|c} 
$L$ & $u_2$ & $\gamma_1$  \\
\hline
4 & $\frac{1}{\sqrt{12}}$ & 6 \\[2 mm]
5 & $\frac{1}{2}$ & 4 \\[2 mm]
6 & $\frac{1}{2} \sqrt{1 \pm \frac{2}{\sqrt{5}}}$ & $5 \mp \sqrt{5}$ \\[2 mm]
7 & $\frac{\sqrt{3}}{2}, \, \frac{1}{\sqrt{12}}$ & 2, \, 6 \\[2 mm]
8 & 1.03826 & 1.50604\\
 & 0.398737 & 4.89008 \\
 & 0.114122 & 7.60388 \\[2 mm]
9 & $\frac{1}{2}(\sqrt{2} \pm 1), \, \frac{1}{2}$ & $4 \mp 2 \, \sqrt{2}, \, 4$
\end{tabular}
\vskip 2 mm
\end{center}
\end{table}

The spin chain model for the $su(2)$ sector has been extended to the more complete set of excitations \cite{Beisert:2005fw}
\beq
\{\Phi^{ab'}, \, \Psi^{\alpha b'}, \bar \Psi^{a \dot \beta}, \, D^{\alpha \dot \beta} \} \,,
\eeq
with $a, \, \alpha, \, \dot \beta \, \in \, \{1,2\}, \  b' \, \in \, \{3,4\}$.
These can be written as a tensor product of two $su(2|2)$ excitations: 
\beq
\chi^{A B'} \eqsp \chi^A \otimes \bar \chi^{B'} \, , \qquad A \eqsp \{a, \alpha\}, \ B' \eqsp \{b', \dot \beta\} \, .
\eeq
We will henceforth refer to the magnons $\chi^A, \bar \chi^{B'}$ as the \emph{left} and the \emph{right chain}, respectively. The phase called $S$ above now becomes a true \emph{matrix}. Its action is given by two identical copies of the same $su(2|2)$ invariant $S$ matrix --- one acting on the $A$ index, and the other one acting on the $B'$ one --- normalised by a common factor $S$ as given in \eqref{eipS}. Further, the effect of higher-loop Feynman diagrams can be included changing to the so-called Zhukowsky variable $x^\pm(u,g^2)$ and introducing a certain phase factor \cite{Beisert:2004hm,Arutyunov:2004vx,Beisert:2006ez}.

Now, the full set of fields of the \nFour model also comprises $\{ \Phi^{12}, \, \Psi^{\alpha a}, \, \bar \Psi^{a' \dot \beta}, F^{\alpha \beta}, \dot{F}^{\dot \alpha \dot \beta} \}$, which apparently do not fit into the $A \, \otimes \, B'$ tensor structure as each of them carries two indices from the same representation. Yet, as mentioned above, the missing fields are all secretly present as double excitations \cite{Eden:2022ipm,Eden:2023ygu}. While they are invisible in the level one wave function, their existence allows us to trust the Bethe equations where we otherwise would not by the quantum numbers --- for instance, how could four fermions fit on a chain of length two without double occupation? 

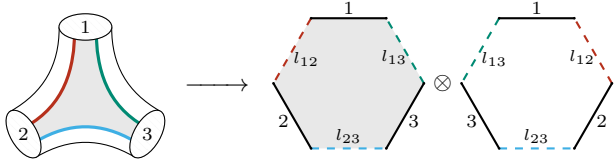
\begin{figure}[htb]
\begin{tikzpicture}
\def\scale{1.0}
\def\scaleII{0.7}
\def\shift{3.5}
\def\shiftI{-3.5}
\def\shiftII{2.5}
\def\colorOne{black}
\def\colorTwo{black}

\def\colorThree{BrickRed}
\def\colorFive{CornflowerBlue}

\def\colorFour{PineGreen}

\coordinate  (A) at ($ ({\scale*cos(0)} , {\scale*sin(0)}) $);
\coordinate (B) at ($ ({\scale*cos(60)} , {\scale*sin(60)}) $); 
\coordinate (C) at ($ ({\scale*cos(120)} , {\scale*sin(120)}) $);
\coordinate (D) at ($ ({\scale*cos(180)} , {\scale*sin(180)}) $);
\coordinate (E) at ($ ({\scale*cos(240)} , {\scale*sin(240)}) $);
\coordinate (F) at ($ ({\scale*cos(300)} , {\scale*sin(300)}) $);


\coordinate  (Ac) at ($ ({(\scale)*cos(330+20)} , {\scale*sin(330+20)-0.2}) $);
\coordinate (Bc) at ($ ({\scale*cos(90-20)} , {\scale*sin(90-20)-0.2}) $); 
\coordinate (Cc) at ($ ({\scale*cos(90+20)} , {\scale*sin(90+20)-0.2}) $);
\coordinate (Dc) at ($ ({\scale*cos(210-20)} , {\scale*sin(210-20)-0.2}) $);
\coordinate (Ec) at ($ ({\scale*cos(210+20)} , {\scale*sin(210+20)-0.2}) $);
\coordinate (Fc) at ($ ({\scale*cos(330-20)} , {\scale*sin(330-20)-0.2}) $);

\coordinate (p1) at ($ ({0.95*\scale*cos(90)+\shiftI} , {0.95*\scale*sin(90)-0.2}) $);
\coordinate (p2) at ($ ({0.95*\scale*cos(210)+\shiftI} , {0.95*\scale*sin(210)-0.2}) $);
\coordinate (p3) at ($ ({0.95*\scale*cos(330)+\shiftI} , {0.95*\scale*sin(330)-0.2}) $);


\draw[thick]($ (Ac)+(\shiftI, 0) $) to[out=150,in=270] 
	($ (Bc)+(\shiftI, 0) $) to[out=140,in=40] 
	 ($ (Cc)+(\shiftI, 0) $) to[out=270,in=30] 
	  ($ (Dc)+(\shiftI, 0) $) to[out=340,in=80] 
	 ($ (Ec)+(\shiftI, 0) $) to[out=30,in=150] 
	 ($ (Fc)+(\shiftI, 0) $) to[out=100,in=200] 
	 ($ (Ac)+(\shiftI, 0) $);

\fill[white]($ (Ac)+(\shiftI, 0) $) to[out=150,in=270] 
	($ (Bc)+(\shiftI, 0) $) to[out=140,in=40] 
	 ($ (Cc)+(\shiftI, 0) $) to[out=270,in=30] 
	  ($ (Dc)+(\shiftI, 0) $) to[out=340,in=80] 
	 ($ (Ec)+(\shiftI, 0) $) to[out=30,in=150] 
	 ($ (Fc)+(\shiftI, 0) $) to[out=100,in=200] 
	 ($ (Ac)+(\shiftI, 0) $);

\fill[white ]($ (Ac)+(\shiftI-0.2, -0.2) $) to[out=150,in=270] 
	 ($ (Bc)+(\shiftI-0.2, 0) $)to[out=140,in=40] 
	 ($ (Cc)+(\shiftI+0.2, 0) $) to[out=270,in=30] 
	  ($ (Dc)+(\shiftI+0.2, -0.2) $)to[out=260,in=160] 
	 ($ (Ec)+(\shiftI,+0.2) $) to[out=30,in=150] 
	 ($ (Fc)+(\shiftI, +0.2) $) to[out=20,in=280]  
	 ($ (Ac)+(\shiftI-0.2, -0.2) $);
\fill[lightgray, opacity=0.35 ]($ (Ac)+(\shiftI-0.2, -0.2) $) to[out=150,in=270] 
	 ($ (Bc)+(\shiftI-0.2, 0) $)to[out=140,in=40] 
	 ($ (Cc)+(\shiftI+0.2, 0) $) to[out=270,in=30] 
	  ($ (Dc)+(\shiftI+0.2, -0.2) $)to[out=260,in=160] 
	 ($ (Ec)+(\shiftI,+0.2) $) to[out=30,in=150] 
	 ($ (Fc)+(\shiftI, +0.2) $) to[out=20,in=280]  
	 ($ (Ac)+(\shiftI-0.2, -0.2) $);

\draw [\colorFour, very thick] ($ (Ac)+(\shiftI-0.2, -0.2) $) to[out=150,in=270] ($ (Bc)+(\shiftI-0.2, 0) $);
\draw [\colorThree, very thick] ($ (Cc)+(\shiftI+0.2, 0) $) to[out=270,in=30]  ($ (Dc)+(\shiftI+0.2, -0.2) $);
\draw [\colorFive, very thick] ($ (Ec)+(\shiftI, +0.2) $) to[out=30,in=150]  ($ (Fc)+(\shiftI, +0.2) $);

\draw[rotate around={90:(p1)},black,  fill=white]   (p1) ellipse (5.0pt and 10pt);
\draw[rotate around={90:(p1)},black, pattern=north west lines , opacity=0.15]   (p1) ellipse (5.0pt and 10pt);

\draw[rotate around={30:(p2)},black, fill=white]   (p2) ellipse (5.0pt and 10pt);
\draw[rotate around={30:(p2)},black, pattern=north west lines , opacity=0.15]   (p2) ellipse (5.0pt and 10pt);

\draw[rotate around={150:(p3)},black, fill=white]   (p3) ellipse (5.0pt and 10pt);
\draw[rotate around={150:(p3)},black, pattern=north west lines , opacity=0.15]  (p3) ellipse (5.0pt and 10pt);

\draw ($ ({0.95*\scale*cos(90) +\shiftI} , {0.95*\scale*sin(90)-0.2}) $)  node {\scriptsize{$1$}} ;
\draw ($ ({0.95*\scale*cos(210)+\shiftI } , {0.95*\scale*sin(210)-0.2}) $)  node {\scriptsize{$2$}} ;
\draw ($ ({0.95*\scale*cos(330) +\shiftI} , {0.95*\scale*sin(330)-0.2}) $)  node {\scriptsize{$3$}} ;


\fill[lightgray, opacity= 0.35]($ (A)+(0, 0) $) to
	 ($ (B)+(0, 0) $)to 
	 ($ (C)+(0, 0) $) to
	  ($ (D)+(0, 0) $)to
	 ($ (E)+(0, 0) $) to
	 ($ (F)+(0, 0) $) to
	 ($ (A)+(0, 0) $);

\draw [\colorFour,  thick, dashed] (A) -- (B);
\draw [\colorTwo, thick] (B) -- (C);
\draw [\colorThree,  thick, dashed] (C) -- (D);
\draw [\colorTwo,  thick] (D) -- (E);
\draw [\colorFive,  thick, dashed] (E) -- (F);
\draw [\colorTwo,  thick] (F) -- (A);

\fill [\colorOne] (A) circle (0.5pt);
\fill [\colorOne] (B) circle (0.5pt);
\fill [\colorOne] (C) circle (0.5pt);
\fill [\colorOne] (D) circle (0.5pt);
\fill [\colorOne] (E) circle (0.5pt);
\fill [\colorOne] (F) circle (0.5pt);

\draw ($ ({1*\scale*cos(90)} , {1*\scale*sin(90)}) $)  node {\scriptsize{$1$}} ;
\draw ($ ({1*\scale*cos(210)} , {1*\scale*sin(210)}) $)  node {\scriptsize{$2$}} ;
\draw ($ ({1*\scale*cos(330)} , {1*\scale*sin(330)}) $)  node {\scriptsize{$3$}} ;


\fill[white]($ (A)+(\shiftII, 0) $) to
	 ($ (B)+(\shiftII, 0) $)to 
	 ($ (C)+(\shiftII, 0) $) to
	  ($ (D)+(\shiftII, 0) $)to
	 ($ (E)+(\shiftII, 0) $) to
	 ($ (F)+(\shiftII, 0) $) to
	 ($ (A)+(\shiftII, 0) $);

\draw [\colorThree,  thick, dashed] ($ (A)+(\shiftII, 0) $) -- ($ (B)+(\shiftII, 0) $) ;
\draw [\colorOne,  thick] ($ (B)+(\shiftII, 0) $) -- ($ (C)+(\shiftII, 0) $);
\draw [\colorFour,   thick, dashed] ($ (C)+(\shiftII, 0) $) -- ($ (D)+(\shiftII, 0) $);
\draw [\colorOne,  thick]  ($ (D)+(\shiftII, 0) $) -- ($ (E)+(\shiftII, 0) $);
\draw [\colorFive,  thick, dashed] ($ (E)+(\shiftII, 0) $) -- ($ (F)+(\shiftII, 0) $);
\draw [\colorOne,  thick] ($ (F)+(\shiftII, 0) $) -- ($ (A)+(\shiftII, 0) $);

\coordinate (l1h1) at ($ ({0.7*\scale*cos(30)} , {0.7*\scale*sin(30)}) $);
\coordinate (l2h1) at ($ ({0.65*\scale*cos(150)} , {0.65*\scale*sin(150)}) $);
\coordinate (l3h1) at ($ ({0.7*\scale*cos(270)} , {0.7*\scale*sin(270)}) $);

\coordinate (l1h2) at ($ ({0.7*\scale*cos(30)+\shiftII} , {0.7*\scale*sin(30)}) $);
\coordinate (l2h2) at ($ ({0.65*\scale*cos(150)+\shiftII} , {0.65*\scale*sin(150)}) $);
\coordinate (l3h2) at ($ ({0.7*\scale*cos(270)+\shiftII} , {0.7*\scale*sin(270)}) $);

\node at (l1h1) { $ {\scriptstyle l_{13}} $ };
\node at (l3h1) { $ {\scriptstyle l_{23}} $ };
\node at (l2h1) { $ {\scriptstyle l_{12}} $ };

\node at (l1h2) { $ {\scriptstyle l_{12}} $ };
\node at (l2h2) { $ {\scriptstyle l_{13}} $ };
\node at (l3h2) { $ {\scriptstyle l_{23}} $ };

\draw ($ ({1*\scale*cos(90) +\shiftII} , {1*\scale*sin(90)}) $)  node {\scriptsize{$1$}} ;
\draw ($ ({1*\scale*cos(210) +\shiftII} , {1*\scale*sin(210)}) $)  node {\scriptsize{$3$}} ;
\draw ($ ({1*\scale*cos(330) +\shiftII} , {1*\scale*sin(330)}) $)  node {\scriptsize{$2$}} ;


\fill [\colorOne] ($ (A)+(\shiftII, 0) $) circle (0.5pt);
\fill [\colorOne] ($ (B)+(\shiftII, 0) $) circle (0.5pt);
\fill [\colorOne] ($ (C)+(\shiftII, 0) $) circle (0.5pt);
\fill [\colorOne] ($ (D)+(\shiftII, 0) $) circle (0.5pt);
\fill [\colorOne] ($ (E)+(\shiftII, 0) $) circle (0.5pt);
\fill [\colorOne] ($ (F)+(\shiftII, 0) $) circle (0.5pt);

\draw  ($ (0,0) + ({\shiftI/2} , 0)$) node {$\xrightarrow{\qquad}$};
\draw  ($ (0,0) + ({\shiftII/2} , 0)$) node {$\otimes$};

\end{tikzpicture}
\caption{Splitting a three-point function into two hexagons.}
\label{fig:hex}
\end{figure}

In Fig. \ref{fig:hex} we depict a three-point function of 
spin chains, or equivalently a three-string vertex. An efficient integrable systems approach to three-point computations has been devised in \cite{Basso:2015zoa}. The three-vertex is split into its back and front surface yielding hexagonal patches. In the figure, the \emph{virtual edges} are coloured. These correspond to bunches of propagators stretching between the operators. The \emph{physical edges} representing spin chains are marked in black. 

The spin chains have to be split as well, taking into account all distributions of excitations or \emph{magnons}:
\begin{equation}
\begin{aligned}
\psi^{l_1 + l_2}_{\{X_1,X_2\}}  \equiv \, & \psi^{l_1}_{\{X_1,X_2\}} \, \psi^{l_2}_{ \{\}} - e^{i \, p_2 \, l_1} \,\psi^{l_1}_{\{X_1\}} \, \psi^{l_2}_{\{X_2\}} \\ & -  e^{i \, p_1 \, l_1} S_{12}  \ \psi^{l_1}_{\{X_2\}} \, \psi^{l_2}_{\{X_1\}} \\ &+ e^{i (p_1+p_2) l_1} \, \psi^{l_1}_{\{\}} \, \psi^{l_2}_{\{X_1,X_2\}} \label{entSU2}
\end{aligned}
\end{equation}
where the symbol $\psi^l_{\{\ldots\}}$ denotes a Bethe wave function of length $l$ with a given set of magnons. Shift operator and $S$~matrix are as in \eqref{eipS}. The split wave function was baptised an \emph{entangled state} in \cite{Escobedo:2010xs}.

Before tackling the correlator $\la \cL \, \cK \, \tilde \cK \ra$ we ultimately want to evaluate, we review the hexagon approach on the example of the simpler but related three-point function $\la \mathbb{1} \, \cK \, \tilde \cK \rangle$, which computes the norm of the operator $\cK$ \cite{Basso:2015zoa}. To decrease the number of terms it is advisable to focus on \emph{transverse excitations}, which do not mix with the vacuum under the \emph{twisted translation} \cite{Basso:2015zoa} used to put the outer points to the standard positions $0,1,\infty$. 

For transverse scalars $X$ on $\cK$ we have the rapidities $u_5=-u_6$, while the operator $\tilde \cK$ must carry the conjugate excitations $\bar X$ with rapidities $u_7=-u_8$. The corresponding spin chain is split as in \eqref{entSU2},
which allows us to express the correlator in terms of hexagons as
\begin{equation}
\begin{aligned}
{\cal A} \,= \, \sum_{\substack{\alpha \, \cup \, \bar \alpha = \{u_5,u_6\} \\ \beta \, \cup \, \bar \beta = \{u_7,u_8\} }}
\begin{tabular}{l}\\$ (-1)^{|\alpha| + |\beta|} \omega(l_{23},\alpha, \bar \alpha) \, \omega(l_{23},\beta, \bar \beta) *$\\$ \mathfrak{h}(\{\},\alpha,\bar \beta) \ \mathfrak{h}(\{\},\beta, \bar \alpha) \,.$\end{tabular}
\end{aligned}
\end{equation}
Here $\alpha,\, \bar \alpha$ and $\beta,\, \bar \beta$ denote the partitions of the sets of magnons into two subsets, and the \emph{splitting factors} $\omega$ can be inferred from \eqref{entSU2}. In the Born approximation we do not require coupling constant corrections to the splitting factors.

In order to evaluate the hexagon form factor we first need to move all magnons to the same physical edge by using \emph{crossing} transformations \cite{Basso:2015zoa}. This sends
\begin{align}
x^\pm \, \xrightarrow{\makebox[0.7cm]{$\pm 2 \gamma$ }} \frac{g^2}{2 \, x^\pm}\, ,
\end{align}
so clearly the operation does not commute over the $g$ expansion. Therefore, the Zhukowsky variables $x^{\pm}$ can only be restricted to the Born approximation after crossing. Second, on the hexagon the $su(2|2)$ invariant $S$ matrix \cite{Beisert:2005tm} is multiplied by the scalar factor $h$ \cite{Basso:2015zoa} that has a monodromy under crossing, since it contains the BES dressing phase \cite{Beisert:2006ez}. 
In the conventions of \cite{Basso:2015zoa} we associate $0 \, \gamma$ crossing to edge 1 of both hexagons, on the back hexagon (white hexagon in Fig. \ref{fig:hex}) we assume $-2 \, \gamma$ for magnons of operator 2 and $-4 \, \gamma$ for magnons of operator 3. On the front (grey) $-2 \, \gamma$ for operator 3 and $-4 \, \gamma$ for operator 2. Note also that crossing sends $X \rar -X$ and $\bar X \rar -\bar X$.

Further, as in the spectral problem the excitations are represented as $\chi^{AB'} \, = \, \chi^A \otimes \bar \chi^{B'}$. Again, we arrange all $\chi$'s on a \emph{left} and all $\bar \chi$'s on a \emph{right chain}. Next, the hexagon scatters only one of the chains using the S matrix \cite{Beisert:2005tm} and finally the left and right chain are contracted employing the rule
\beq
\langle \phi^a_j | \bar \phi^{b'}_j \rangle \, = \, \epsilon^{ab'} \, , \qquad \langle \psi^\alpha_j | \bar \psi^{\dot \beta}_j \rangle \, = \, \epsilon^{\alpha \dot \beta} \, . \label{hexaContract}
\eeq
This assignment is valid in the \emph{spin chain frame} \cite{Basso:2015zoa}\footnote{In the \emph{string frame} the rhs. of the second rule in \eqref{hexaContract}  would pick up a factor $i$.}.

In the norm computation, only hexagons with equal numbers of $X$ and $\bar X$ magnons can be non-vanishing because the transverse scalars cannot be contracted on the twisted vacuum. The only distinct amplitudes are
\begin{equation}
\begin{aligned}
\la \mh | \{ \} \{ \} \{ \} \ra  = &  1 \, , \\
\la \mh | \{ \} \{ X_5 \} \{\bar X_7\} \ra = & \frac{i}{u_5-u_7} e^{i \, p_7} \, , \label{xNormAmps} \\
\la \mh | \{ \} \{ X_5 X_6 \} \{\bar{X}_7 \bar{X}_8 \} \ra  = &
  \frac{4 \, u_5 \, u_7 \cdot e^{2 \, i \, p_5} }{(2 \,  u_5 + i)(2 \, u_7 + i)} \, *\\&\frac{2 \, u_5^2 + 2 \, u_7^2 + 1}{(u_5-u_7)^2 (u_5 + u_7)^2 }
\end{aligned}
\end{equation}
In the third row of the last equation we have identified $u_6 \eqsp - u_5, \, u_8 \eqsp - u_7$ to obtain a concise formula. Note that both non-trivial amplitudes contain \emph{particle creation poles} that occur when conjugate magnons occupy different physical edges. 
Since both $\cK$ and $\tilde \cK$ are characterised by the same rapidities we cannot identify $u_7 = u_5$ without factoring out or regularising \footnote{In \cite{Eden:2022djp,Eden:2022ipm} it was attempted to introduce twist into the Bethe equations.} the real poles. Fortunately, they do factor out upon adding up all the partitions.
Yet, this can only happen upon omitting the explicit momentum factors from equation \eqref{xNormAmps}. The need to drop these can be demonstrated deriving the amplitudes in the string frame and switching to the spin chain picture employing the formulae in \cite{Basso:2015zoa,Caetano:2016keh}.
Here we use the rule
\beq
2 \, u - i \rar e^{-i \, p} \, (2 \, u + i) 
\eeq
since denominator factors of the form $(2 \, u_5 + i)$ can eg. arise from the difference $x^+_5 - x^-_6$ in $S$ matrix elements. The need to rescale by momentum factors has been realised in the original paper \cite{Basso:2015zoa} and was extended for fermions in \cite{Caetano:2016keh}.

After factorisation we can identify $u_7 \eqsp u_5$. By way of example, the length 4 and 5 results are
\begin{equation}
\begin{aligned}
\la \mathbb{1} \ \cK^4 \, \tilde \cK^4 \ra = & \frac{512 \, u_5^2 \, e^{i \, p_5}}{(1 + 4 \, u_5^2)^7} \, (23 - 128 \, u_5^2 + 992 \, u_5^4\\
&\hspace{1.5cm}- 512 u_5^6 + 768 \, u_5^8) \, , \label{fewNorms} \\
\la \mathbb{1} \ \cK^5 \, \tilde \cK^5 \ra = & \frac{512 \, u^5_2 \, e^{i \, p_5}}{(1 + 4 \, u_5^2)^9} \, (65 - 952 \, u_5^2 + 12336 \, u_5^4\\
&\hspace{1.5cm}- 38144 \, u_5^6 + 87808 \, u_5^8\\
&\hspace{1.5cm} - 30720 \, u_5^{10} + 20480 \, u_5^{12}) \, .
\end{aligned}
\end{equation}
Substituting the rapidities for $L \eqsp 4,5$ from Tab.~\ref{tab:Tab1} we find the values $108 \, (-1)^{4/3}, \, -80 \, i$, respectively, confirming \cite{Basso:2015zoa,Eden:2016xvg}
\beq
\la\mathbb{1} \, \cK \, \tilde \cK \ra \eqsp {\cal G} \, S_{12}
\eeq
where ${\cal G}$ is the \emph{Gaudin norm} \cite{Gaudin:1996}. \\

The $\la \cL \, \cK \, \tilde \cK \ra$ computation additionally requires an entangled state at point 1 distributing the four magnons of $\cL$ with rapidities $u_1,\dots, u_4$ over the two hexagons. This considerably augments the number of possible hexagon amplitudes. In the Born approximation there is no $e^{i \, p}$ rescaling for the magnons of $\cL$ because of their vanishing momenta (viz infinite rapidities). Thus the rescaling is tied solely to the occupation by scalar excitations from $\cK, \, \tilde \cK$ at given momentum and crossing. We can infer it from \eqref{xNormAmps} where an inverse image in $\la \mathbb{1} \, \cK \, \tilde \cK \ra$ exists. However, a new feature of $\la \cL \, \cK \, \tilde \cK \ra$ is the occurence of non-vanishing hexagon amplitudes with one or three of the scalars from $\cK, \tilde \cK$ and also one or three of the magnons of $\cL$.
Fortunately the rescaling should not depend on the flavour of the scalars in question so that we can compare to a norm calculation with longitudinal scalars where necessary.
In Tab.~\ref{tab:Tab2} we list the additional hexagon amplitudes \footnote{The longitudinal scalar $Y$ is self-conjugate on the hexagon.} and the momentum factors which need to be removed.
\begin{table}[h]
\caption{Additional momentum factors for amplitudes containing magnons from $\cK, \, \tilde \cK$.} \label{tab:Tab2} \vskip 2mm
\begin{tabular}{c|c} 
hexagon & factor \\
\hline
$\la \mh | \{ \ldots \} \{ Y_5\} \{ \} \ra, \, \la \mh | \{ \ldots \} \{ \} \{ Y_7 \} \ra^{\phantom{\dot \beta}}$ & 1 \\[2 mm]
$\la \mh | \{ \ldots \} \{ Y_5 Y_6\} \{ \} \ra$ & $e^{2 \, i \, p_5}$ \\[2 mm]
$\la \mh | \{ \ldots \} \{ Y_5 Y_6\} \{ Y_7\} \ra$ & $e^{2 \, i \, p_5} \, e^{2 \, i \, p_7}$ \\[2 mm]
$\la \mh | \{ \ldots \} \{ \} \{ Y_7 Y_8\} \ra, \,  \la \mh | \{ \ldots \} \{ Y_5 \} \{ Y_7 Y_8\}$ & 1 
\end{tabular}
\end{table}

According to \eqref{q4L}, the Lagrange operator is a four-fold supersymmetry descendent of the vacuum. We therefore expect four infinite Bethe roots. 
In order to find a regulated Bethe solution for the Yang-Mills term of the Lagrange operator as derived in \cite{Eden:2023ygu} a twist regulator in a nested Bethe-ansatz for an $su(2|2)$ sector was employed. 
To this end a factor $e^{i \, m_j \, \beta}$ is introduced into the level-$j$ Bethe equation, with a universal order parameter $\beta \, \ll \, 1$. The Bethe roots are assumed to have an expansion
\beq
 u_j \rar \frac{u_{j,-1}}{\beta} + u_{j,0} + u_{j,1} \, \beta + \ldots
\eeq
of which we will only need the leading order in $\beta$ in this letter. The hexagon calculation only depends on the level one roots, so we just need
\beq
u_{j,-1} \in \left\{ \frac{+ 1 + i}{\sqrt{2}},\,\frac{+ 1 - i}{\sqrt{2}},\,\frac{- 1 + i}{\sqrt{2}},\,\frac{-1 - i}{\sqrt{2}} \right\}
\eeq
with $j=1,\dots,4$. We will henceforth drop the -1 label on these roots. Note that they satisfy $\prod_j u_j \eqsp 1$.

Since twist is needed on the supersymmetry and on an auxiliary $R$ symmetry node one might wonder whether supersymmetry and $R$ invariance are manifest in every partial wave function of the multi-component ansatz or rather only in the sum over all 36 parts. For the Yang-Mills term of the Lagrangian the individual wave functions are degenerate --- will we find the same in our three-point calculation?

To form the entangled state of $\mathcal{L}$ we generalise \eqref{entSU2} by scattering the four fermions with the full $su(2|2)^2$ $S$~matrix~\cite{Beisert:2005tm}. \emph{Born level} means restricting to leading order in $g$ and $\beta$. It is advisable to do so from the very beginning, resulting in the simple scattering
\begin{equation}
\begin{aligned}
\phi^a_1 \, \psi^\gamma_2 & \rightarrow \psi^\gamma_2 \, \phi^a_1  \\
\psi^\alpha_1 \, \phi^c_2 & \rightarrow \phi^c_2 \, \psi^\alpha_1\\
\phi^a_1 \, \phi^c_2 & \rightarrow  \phi^c_2 \, \phi^a_1 + \frac{\tilde g}{u_1 \, u_2} \ep^{ac} \ep_{\alpha \gamma}  \, \psi^\alpha_2 \, \psi^\gamma_1 \, \cZ^-   \\
\psi^\alpha_1 \, \psi^\gamma_2 & \rightarrow  \psi^\gamma_2 \, \psi^\alpha_1 + \frac{\tilde g}{u_1 \, u_2} \ep^{\alpha \gamma} \ep_{a c}  \,  \phi^a_2 \, \phi^c_1 \, \cZ^+ \label{scatterState}
\end{aligned}
\end{equation}
on both chains. Here we introduced $\tilde g \eqsp (g \, \beta^2)/\sqrt{2}$. Thus scattering is diagonal up to the remnants of the $C_{12}$ and $F_{12}$ elements of the $S$ matrix \cite{Beisert:2005tm}, where the $\mathcal{Z}^{\pm}$ indicate length changing effects. In \cite{Eden:2023ygu} the calculation
\beq
\la \cL \ \cO^L \, \cO^L \ra \eqsp 0
\eeq
with two half-BPS states $\cO$ of length $L$ was presented. This computation is chiral in the sense that four fermions can only be self-contracted under certain conditions. Namely, the scattering processes to form the entangled state as well as those in the evaluation of the hexagons need to produce in total $C^2, \, C \, F, \, F^2$ (or higher) for every term. Thereby, all the leading contributions come with the factor
\beq
\frac{\tilde g^2}{\prod_j u_j} \rar \tilde g^2
\eeq
if we go on shell. In fact, the actual Bethe solution is never needed because the rapidity dependence is global. Furthermore, the length-changing effects denoted by $\mathcal{Z}^{\pm}$ referring to the insertion/deletion of a site in \eqref{scatterState} are irrelevant because all magnons have infinite rapidities.

On the other hand, when evaluating $\la \cL \ \cK \, \tilde \cK \ra$ length changing is of the utmost importance since there are magnons with finite rapidities. 
Recall that we can scatter on either chain when computing hexagons. For instance, scattering on the right chain produces $C$ elements so that only spin chain shortening operations $\cZ^-$ appear. We can then move the $Z$~markers to the left using \cite{Beisert:2005tm}
\beq
\chi^{B'} \, \cZ^\pm \rar e^{\pm i \, p} \, \cZ^\pm \, \chi^{B'} \, ,
\eeq
and remove them from the form factor as  \cite{Basso:2015zoa}
\beq
\begin{aligned}
\la \mathfrak{h} | &\cZ^\pm \, \chi^{B_1'}(p_1) \ldots \chi^{B_n'}(p_n) \ra \eqsp \\ &\prod_{j = 1}^n e^{\mp i \, p_j /2} \, \la \mathfrak{h} | \chi^{B_1'}(p_1) \ldots \chi^{B_n'}(p_n) \ra \, . \label{bkvDoZ}
\end{aligned}
\eeq
Note that $\pm2 \, \gamma$ on the second edge inverts momentum factors. Further, crossing can create entire powers of $e^{i \, p}$ when acting on the Zhukowsky variables in the usual combinations like $x^+ - y^-, \, 1 - g^2/(2 \, x^- y^+), \, \ldots$ in the $S$ matrix elements. Half integer powers may arise from the relative normalisation between bosons and fermions:
\begin{equation}
\sqrt{i \, (x^- - x^+)} \bigl|_{u^{\mp 2 \gamma}} = \mp e^{i \, p/2} \, \frac{i \, \sqrt{i \, (x^- - x^+)}}{x^+} 
\biggl|_{u^{0 \, \gamma}}\,
\end{equation}
as well as from \eqref{bkvDoZ}. Finally we have to deal with the $Z$ markers arising from scattering the four fermions according to \eqref{scatterState} when building the entangled state of $\mathcal{L}$. As the Lagrangian is inserted at point~1 and its magnons have vanishing momentum, we can move these markers to the very left without picking up any additional factors. Then we utilise \eqref{bkvDoZ} once again. Using our assumptions on the rapidities (vanishing momenta $p_1 \ldots p_4$ for the magnons of $\cL$ and $u_6 \eqsp - u_5, \, u_8 \eqsp - u_7$ on $\cK$, $\tilde\cK$), we tabulate the extra momentum factors for all contributing partitions and notice that they are equal on the front and back hexagon. We are thus free to choose on which hexagon to act with the markers from the entangled state for $\cL$. Applying them to both would overcount their effect. In a tessellation \cite{Eden:2016xvg,Fleury:2016ykk} with more hexagon tiles around the Lagrangian insertion one will presumably need some averaging prescription \footnote{We are grateful to P.~Vieira for a discussion on this point.}. Last, for these three-point functions a curious observation with respect to the total $Z$ marker action is the symmetry under $\cZ^+ \, \leftrightarrow \, \cZ^-$. This can be traced to the similarity of the $C$ and $F$ elements.

Given all these precautions the miracle happens: there are only integer powers of $e^{i \, p}$ and the particle creation poles factor out as in the norm computation! Despite the fact that 2286 of 35664 possible hexagon amplitudes are non-trivial, there are maximally 180 contributing partitions in any partial wave function so that factorisation can again be achieved analytically. More precisely, the 24 permutations of four distinct fermions \eqref{myFour} each yield 104 partitions, and the results are equal up to the obvious signs. The 12 cases with degenerate fermions \eqref{degenerate} all non-trivially vanish. Here 144, 168 or 180 partitions sum to zero.

In $\la \mathbb{1} \, \cK^L \, \tilde \cK^L \ra$ the width of the edge connecting the operators at points 2, 3 is unambiguously $l_{23} \eqsp L$ because the identity has length 0 \cite{Basso:2015zoa,Eden:2017ozn}. In general, the bridge length reads $l_{23} \, = \, (L_2 + L_3 - L_1)/2$ and for $\la \cL \, \cK^L \, \tilde \cK^L \ra$ we obtain $l_{23} = L - L_1/2$. The leading $\tr(F^2)$ term surely has $L_1 \eqsp 2$. In our factorisation exercise there are always two $Z$ markers in any term.
To the combinations $\cZ^- \cZ^-, \, \cZ^+ \cZ^-, \, \cZ^+ \cZ^+$ we could assign operator length 0, 2, 4, respectively. If we do not want to destroy the cancellation of the particle creation pole, every partition will have to have the same starting value for $l_{23}$ while we rely on the markers to correctly incorporate the length changing effects.

In analogy to \eqref{fewNorms}, for $l_{23} \eqsp 2 \ldots 8$ we find the rational functions given in Tab. \ref{tab:l23func} for the initial ordering $\{\psi^{13}, \, \psi^{14}, \, \psi^{23}, \psi^{24}\}$ of the $\cL$ excitations.

\begin{table}[htb]
\caption{Rational functions obtained from the hexagon evaluation for bridge lengths $l_{23}$.}\label{tab:l23func} \vskip 2mm
\begin{tabular}{l|l}
$l_{23}$& \ rational function\\\hline
2&0   \\
3&$\frac{8192 \, u_5^2 \, e^{i \, p_5}}{(1 + 4 \, u_5^2)^5}$  \\
4&$\frac{8192 \, u_5^2 \, e^{i \, p_5}}{(1 + 4 \, u_5^2)^7} \cdot (1 - 12 \, u_5^2)^2$  \\
5&$\frac{8192 \, u_5^2 \, e^{i \, p_5}}{(1 + 4 \, u_5^2)^9} \cdot 8 \, (1 - 4 \, u_5^2)^2 (1 - 8 \, u_5^2 + 80 \, u_5^4)$ \\
6&$\frac{8192 \, u_5^2 \, e^{i \, p_5}}{(1 + 4 \, u_5^2)^{11}} \cdot (1 - 40 \, u_5^2 + 80 \, u_5^4)^2 (5 - 24 \, u_5^2 + 80 \, u_5^4)$  \\
7&$\frac{8192 \, u_5^2 \, e^{i \, p_5}}{(1 + 4 \, u_5^2)^{13}} \cdot \frac{(3 - 4 \, u_5^2)^2 (1 - 12 \, u_5^2)^2 }{ (3 - 80 \, u_5^2 + 1824 \, u_5^4 - 5376 \, u_5^6 + 8960 \, u_5^8)^{-1}}$   \\
8&$\frac{8192 \, u_5^2 \, e^{i \, p_5}}{(1 + 4 \, u_5^2)^{15}} \cdot \frac{2 \, (1 - 84 \, u_5^2 + 560 \, u_5^4 - 448 \, u_5^6)^2}{(7 - 112 \, u_5^2 + 928 \, u_5^4 - 1792 \, u_5^6 + 1792 \, u_5^8)^{-1}} $
\end{tabular}
\end{table}

Curiously, $l_{23} \eqsp L$ yields zero from the third row on upon substituting the Bethe roots from Tab.~\ref{tab:Tab1}. This confirms operator length two for the full Lagrangian, and indeed for all operators \footnote{At $L \eqsp 8$ we have numerically checked the equivalence up to 200 digits.} in Tab.~\ref{tab:Tab1}
\beq
\frac{\la \cL \, \cK^L \, \tilde \cK^L \ra}{\la \mathbb{1} \ \cK^L \, \tilde \cK^L \ra} \eqsp \frac{\tilde g^2 \, \gamma_1}{L} \label{mainRes}
\eeq
 with $l_{23} \eqsp L-1$. This is our main result.

 \section{Conclusions}
 
We have successfully generalised the simple test in \cite{Eden:2023ygu} to the formula $\la \cL \, \cK \, \tilde \cK \ra \, \propto \, \gamma_1$ as expected from field theory, see \eqref{gamRel}. However, we do not fully understand the normalisation. For a \emph{connected} correlator, the hexagon result should be scaled up by a factor
\beq
\frac{\sqrt{L}}{\sqrt{{\cal G} \, \prod_{i<j} S_{ij}}}
\eeq
per operator. Now, $\la \mathbb{1} \, \cK \, \tilde \cK \ra$ is disconnected, by which token the factor $\sqrt{L}$ should be omitted there, explaining the appearance of the explicit $1/L$ in \eqref{mainRes}. 

How about the normalisation of $\cL$, though? The Gaudin determinant from the full set of Bethe equations including secondary roots equals $144^2 \, \beta^{16}$, whereas the phase $\prod_{i<j} S_{ij}$ associated to the Lagrangian is 1. On the other hand, reading off coefficients for the individual wave functions of the multi-component ansatz from an equivalent nested Bethe ansatz leads to the conjecture that each such coefficient is given by the complete \emph{above level-one wave function} \cite{Eden:2022ipm,Eden:2023ygu}. We have verified this with $L \, > \, 2$ for the off-shell problem with four level-one $Q_{3}^{2}$ excitations and two auxiliary $sl(2)$ (left) and $su(2)$ (right) excitations each.
For example, the coefficient of the $\{\psi^{13}, \, \psi^{14}, \, \psi^{23}, \, \psi^{24}\}$ partial wave function discussed above is the nested wave function for auxiliary excitations with rapidities $\{v_1 w_1, v_2, w_2, 1\}$, respectively, on the four level-one vacuum sites $\psi^{24}(u_i)$. Here we have to scatter the two left wing and the two right wing magnons over each other resulting in four similar blocks. Given the degeneracy of the 24 + 12 wave functions described in the last section we compute an extra factor of $-24 \, \sqrt{3} \, \beta^4$ from the sum of these coefficients put on shell. For the explicit Bethe solution see \cite{Eden:2023ygu}. The right hand side of formula \eqref{mainRes} should thus be rescaled as
\beq
\frac{\beta^4}{2} \, \frac{\gamma_1}{L} * \frac{-24 \, \sqrt{3} \, \beta^4}{144 \ \beta^8} * \sqrt{2 \, L L} \eqsp -\frac{\gamma_1}{\sqrt{4!}} \, .
\eeq
While the powers of $\beta$ fall into place and we may dismiss the minus sign as somewhat accidental, 
we have no good explanation for the factor $\sqrt{4!}$ in the normalisation of the Lagrange operator. Note however, that differences between operator norms in integrability and field theory, respectively, are not uncommon \cite{Basso:2015zoa}.
 
Nevertheless, we have presented evidence for the Lagrangian insertion to work in integrability as in field theory. Our computation is perfectly stable although descendent operators can be quite intricate to handle in the hexagon approach. In fact, explicit knowledge of a Bethe solution for $\cL$ is not required unless we are interested in the exact normalisation in \eqref{gamRel}. If not, we may pick any of the wave functions without degenerate fermions and ignore the rest of the ensemble.

The obvious future applications of the technique would be an extension to other types of operators carrying anomalous dimension, and most of all non-planar corrections to the anomalous dimension of transverse $su(2)$ sector operators by using a tessellation as in \cite{Eden:2017ozn,Bargheer:2017nne,Eden:2022djp}.

\begin{acknowledgments}

B.~Eden is supported by Heisenberg funding of the Deutsche Forschungsgemeinschaft (DFG), grant 441791296 or Ed 78/7-1, M.~Gottwald by the DFG project grant 441793388 or Ed 78/8-1.
D.~le Plat~acknowledges support from the Stiftung der Deutschen Wirtschaft.

\end{acknowledgments}

\appendix

\bibliographystyle{apsrev4-1} 
\bibliography{apssamp}

\end{document}